\documentclass[aps,preprint]{revtex4}
\usepackage{amsfonts}
\usepackage{amsmath}
\usepackage{amssymb}
\usepackage{graphicx}
\usepackage{grffile}
\usepackage{epstopdf}
\usepackage{float}
\usepackage{ulem}
\usepackage{cancel}
\usepackage{xcolor}
\usepackage{cleveref}
\newcommand{\myref}[1]{Eq. \eqref{#1}}
\newcommand{\figref}[1]{FIG. \ref{#1}}

\usepackage{bbold}
\usepackage{bm}
\usepackage{subfigure}
\usepackage{caption}

\begin{document}
	\title{Black Holes in a Cavity: Heat engine and Joule-Thomson Expansion}
	\author{Yihe Cao}
	\email{yihe@stu.scu.edu.cn}
	\author{Hanwen Feng}
	\email{fenghanwen@stu.scu.edu.cn}
	\author{Jun Tao}
	\email{taojun@scu.edu.cn}
	\author{Yadong Xue}
	\email{xueyadong@stu.scu.edu.cn}
	\affiliation{Center for Theoretical Physics, College of Physics, Sichuan University, Chengdu, 610065, China}

\begin{abstract}
We consider the charged d-dimensional black holes in a cavity in extended phase space and investigate the heat engine and the Joule-Thomson (JT) expansion. Since the phase structure of black holes in a cavity is similar to anti-de-sitter (AdS) cases, we take black holes in a cavity as the working substance in the heat engine and calculate their efficiency in Carnot cycle and rectangular cycle. Also, we discuss whether the JT expansion of charged black holes in a cavity are consistent with AdS cases and conclude the effect of different boundary conditions on black hole thermodynamics.
\par 
\end{abstract}
\maketitle

\clearpage

\section{Introduction}
The thermal quantum radiance of black hole reveals the connection among gravity, quantum theory and black hole thermodynamics \cite{York:1983zb,Hawking:1975vcx,Bekenstein:1975tw}.
The idea to place black holes in a cavity is proposed to solve the problem of investigating thermodynamically unstable black hole in asymptotically flat space, and it attracts interests in the study of
 black hole thermodynamics and quantum gravity \cite{York:1986it}. The temperature of the surface of a cavity is fixed with a heat bath surrounded. According to the semi-classical asymptotic relationship between the on-shell Euclidean action and the partition function, we can obtain the expression of free energy. Furthermore, we define the volume of cavity as the one of the system. In this framework, the phase transition, the thermodynamic geometry and the critical behaviour of vary BHs in a cavity are discussed in \cite{Braden:1990hw,Wang:2020hjw,Wang:2019urm,Wang:2019kxp,Wang:2019cax,Wang:2020osg,Zhao:2020nrx,Simovic:2019zgb,Dias:2018zjg}. Another boundary condition to make the black hole thermodynamically stable is to put it in the anti-de sitter(AdS) space, and related work on black hole thermodynamics are studied in \cite{Caldarelli:1999xj,Padmanabhan:2002sha,Kastor:2009wy,Hawking:1982dh,Chamblin:1999tk,Chamblin:1999hg,Kubiznak:2012wp,He:2016fiz}. These results shows the thermodynamic behaviours of AdS-BHs are similar to those of BHs in cavities \cite{Wang:2020hjw}.
 
 Due to this fact, we can investigate its properties as a heat engine like AdS-BH, which is to define a cycle in the extended thermodynamical space and the AdS black holes are considered as the working substance, and the related work can be found in \cite{Mo:2018hav,Rajani:2019ovp,Debnath:2020zdv,Johnson:2019ayc,Balart:2019uok,Yerra:2018mni,Ghaffarnejad:2018gtj,Zhang:2018vqs,Zhang:2018hms,Yerra:2020bfx,Fernando:2018fpq,Hendi:2017bys,Chakraborty:2017weq,Chakraborty:2016ssb,Chandrasekhar:2016lbd,Chandrasekhar:2017zyb}. Moreover, a cavity has explicit boundaries, which makes the correspondence between the classic heat engine and the black hole one more reasonable. We considered charged black holes in a cavity as the heat engine which works in the Carnot cycle and rectangular cycle. The Carnot cycle consists of two isothermal paths and two adiabatic paths, it is always theoretically highest according to the second law of thermodynamics, and thus it provides an upper bound for us to check the calculation. The rectangular cycle is made up of two isochoric paths and two isobaric paths, which can be taken as the smallest unit of a more complicated cycle  \cite{Chakraborty:2016ssb}.
 
As an important property of van der Waals system, the Joule-Thomson (JT) expansion is introduced to AdS-BHs and firstly investigated in \cite{Okcu:2016tgt}, which shows both cooling and heating states of the RN-AdS black hole exist during the isenthalpic process. Furthermore, it's found that the JT expansion for different black holes were found have similarities in isenthalpic curves and inversion curves \cite{Cao:2021dcq,Bi:2020vcg,Guo:2020qxy,Hegde:2020xlv,Rajani:2020mdw,Rostami:2019ivr,Nam:2019zyk,MahdavianYekta:2019dwf,Lan:2018nnp,Ghaffarnejad:2018exz,Pu:2019bxf,Mo:2018qkt,Cisterna:2018jqg,Li:2019jcd,RizwanCL:2018cyb,Chabab:2018zix,Mo:2018rgq,Okcu:2017qgo,Haldar:2018cks,Feng:2020swq,Nam:2020gud,Meng:2020csd,Zhang:2021raw,Liang:2021elg,Liang:2021xny,Graca:2021izb,Mirza:2021kvi,Yin:2021akt,Zhang:2021kha,Biswas:2021uop,Graca:2021ker}, which means the isenthalpic curves divides the heating and the cooling regions and the extreme points are coincident with the inversion curves. Moreover, \cite{Abdusattar:2021wfv} shows different types of isenthalpic curves when it comes to study the JT expansion of the FRW universe. In this paper, we prove the JT expansion of black holes in a cavity does exist, but differ from AdS-BHs. We find the JT coefficient always keeps positive, which means the temperature decrease during this kind of adiabatic expansion. In other words, the black hole can only become colder after the isenthalpic process. And it is different from the situations ever investigated before. 
\par In section \ref{sec:2}, we derive the thermodynamic quantities
of the d-dimensional charged BH in the cavities. We calculate the heat engine efficiency of 4-dimensional Schwarzchild black holes and RN black holes in a cavity in section \ref{sec:HE}. In section
\ref{sec:JT}, the JT expansion of the RN black holes and the higher dimensional cases in a cavity are investigated.  We summarize our results in section \ref{sec:Discussion-and-Conclusions}.

\section{Black holes in a cavity}
\label{sec:2}
The $d$-dimensional Einstein-Maxwell action according to the ref \cite{York:1986lje,Lundgren:2006kt} is given by
\begin{equation}
	\begin{aligned}
			I=I_{bulk}+I_{surf}=&-\frac{1}{16\pi}\int_{\mathcal{M}} d^{d}x\sqrt{-g}R
		+\frac{1}{8\pi}\int_{\partial\mathcal{M}} d^{d-1}x\sqrt{-\gamma}(K-K_{0})	\\&+\frac{1}{16\pi}\int_{\mathcal{M}} d^{d}x\sqrt{-g}F^{2}+\frac{1}{16\pi}\int_{\partial\mathcal{M}} d^{d-1}x\sqrt{-\gamma}n_{\nu}F^{\mu\nu}A_{\mu},
	\end{aligned}
\label{I}
\end{equation}
where $n_{\nu}$ is the unit normal vector of the boundary, $\gamma$ is the metric on the boundary, $K$ is the trace of the extrinsic curvature, and $K_{0}$ is a subtraction term to ensure the Gibbons-Hawking-York term vanish in flat space-time.
The metric of the spherical d-dimensional charged BH reads \cite{Chabab:2016cem}
\begin{equation}
	ds^2=-f(r)dt^2+\frac{1}{f(r)}dr^2+r^2d\omega_{d-2}^2, \thickspace \thickspace \thickspace  A_{t}dt=-\sqrt{\frac{d-2}{2d-6}}\frac{Q}{r^{d-3}}dt.
	\label{ds}
\end{equation}
By varying the action Eq.(\ref{I}), we have the metric function as
\begin{equation}
	f(r)=1+\frac{Q^{2}}{r^{2(d-3)}}-\frac{Q^{2}}{r^{d-3}r_{+}^{d-3}}-\frac{r_{+}^{d-3}}{r^{d-3}},
\end{equation}
where the radius of the event horizon is defined by $f(r_{+})=0$ and the parameter $Q$ is the charge of the black hole.
Corresponding to the Hawking temperature $T_{H}$
by $T=\frac{T_{H}}{\sqrt{f(r_{B})}}$ in \cite{Braden:1990hw}, the temperature of a cavity is defined as
\begin{equation}
	T=\frac{(d-3)\left(1-Q^{2}\text{\ensuremath{\text{\ensuremath{r_{+}}}^{6-2d}}}\right)}{4\pi\text{\ensuremath{r_{+}}}\sqrt{f(r_{B})}},\label{eq:Th}
\end{equation}
where $r_{B}$ is the radius of a cavity.
Furthermore, the Euclidean action corresponds to the action Eq.(\ref{I}) by $I_{E}=iI$. The Euclidean time $\tau$ is related to $t$ by $\tau =it$ and the metric becomes positive infinite \cite{York:1986it,Wang:2019kxp}. In this way, we derive the Euclidean action by substituting Euclidean metric into the action, which gives
\begin{equation}
	I_{E}=\frac{d-2}{8\pi}\frac{\omega_{d-2}r_{B}^{d-3}}{T}(1-\sqrt{f(r_{B})})-S,
\end{equation}
where $S=\frac{\omega_{d-2}}{4}r_{+}^{d-2}$ is the entropy of the
black hole. $\omega_{d}=\frac{2\pi^{\frac{d+1}{2}}}{\Gamma(\frac{d+1}{2})}$ is the volume of the unit $d$-sphere where $\Gamma$ represents the gamma function. The free energy is related to the Euclidean action in an approximately semi-classical way as $F=-TlnZ=TI_{E}$ \cite{Wang:2019urm}, and thus the thermal energy of the black hole in a cavity is $E=-T^2\frac{\partial F/T}{\partial T}$, which yields as
\begin{equation}
	E=\frac{(d-2)r_{B}^{d-3}\omega_{d-2}\left(1-\sqrt{f(r_{B})}\right)}{8\pi}.
\end{equation}
The thermodynamic volume $V$ of the system is defined as the volume of the cavity
\begin{equation}
	V\equiv\frac{r_{B}^{d-1}\omega_{d-2}}{d-1},
\end{equation}
and the conjugate thermodynamic pressure naturally arises according to $P=-\frac{\partial E}{\partial V}$.
In this new extended phase space, the enthalpy $H$ can be derived by $H=E+PV$.

\section{Black holes as heat engines in a cavity}
\label{sec:HE} 
For the purpose of comparing the heat engine properties of black hole in a cavity with AdS-BH, we consider a cavity as the container and take black holes as the working substance in the pressure-volume space.
\subsection{Schwarzchild black holes in a cavity}
\label{sec:sch}
In the following sections, the black hole is set in thermodynamical cycles. For simplicity, we focus on the case of $d=4$. Four-dimentional space-time metric of RN black hole yields
\begin{equation}
	ds^{2}=-f(r)dt^{2}+\frac{dr^{2}}{f(r)}+r^{2}(d\theta^{2}+\sin^{2}\theta d\phi^{2}),
\end{equation}
where $\ensuremath{f(r)}$ is expressed as  
\begin{equation}
	f(r)=(1-\frac{r_{+}}{r})(1-\frac{Q_{b}^{2}}{r_{+}r}),
	\label{eq:RNf-1-1}
\end{equation} and the effective potential satisfies $A=A_{t}dt=-\frac{Q_{b}}{r}dt$. The event horizon radius is represented by $r_{+}$, meanwhile the charge of the black hole is represented by $Q_{b}.$ Subsequently the temperature of this system with radius $r=r_{B}>r_{+}$ had a definition in \cite{Braden:1990hw},
and for $d=4$, 
\begin{equation}
	T(r_{B},x,Q)=\frac{T_{H}}{\sqrt{f(r_{B})}}=\frac{1-\frac{Q^{2}}{r_{B}^{2}x^{2}}}{4\pi r_{B}x\sqrt{f(r_{B})}},\label{eq:T4}
\end{equation}
where $Q_{b}=Q$, $r_{B}$ is the radius of a cavity and $T_{H}=\frac{1}{4\pi} f'(r_{+})$ is the Hawking temperature of the black hole. It is worth noting that the
physical range of the event horizon is constrained as $\text{\ensuremath{\frac{r_{e}}{r_{B}}}}\leq x\text{\ensuremath{\equiv}}\frac{r_{+}}{r_{B}}\leq1$
in which $r_{e}=Q$ is the horizon radius of the extremal RN-BH. 
\par First, we start from the Schwarzchild black holes, which is the form of RN black holes reduce to when $Q=0$. Under these conditions, \myref{eq:T4} takes the form as
\begin{eqnarray}
	T(r_{B},x)=\frac{T_{H}}{\sqrt{f(r_{B})}}=\frac{1}{4\pi r_{B}x\sqrt{1-x}}.
	\label{scht}
\end{eqnarray}
and the conjugate pressure is
\begin{eqnarray}
	P(r_{B},x)=\frac{2r_{B}^{2}x-r_{B}^{2}x^{2}}{8\pi r_{B}^{4}x\sqrt{1-x}}-\frac{1}{4\pi r_{B}^{2}}.
	\label{schp}
\end{eqnarray}
We can derive the expression for $x$ in terms of $r_{B}$ and $P$ with \myref{schp}, 
\begin{eqnarray}
	x=4 \left(\left(4 \pi  P r_{B}^3+r_{B}\right)\sqrt{2 \pi P \left(2 \pi  P r_{B}^2+1\right)}-4 \pi  P r_{B}^2 \left(2 \pi  P r_{B}^2+1\right)\right).
	\label{x}
\end{eqnarray}
With this result, the temperature can be rewritten as
\begin{gather}
	T(r_{B}, P)=(16 \pi  r_{B} \Gamma\sqrt{1+4\Gamma})^{-1}, \\
	\Gamma=\sqrt{2 \pi } \sqrt{P \left(2 \pi  P r_{B}^2+1\right) \left(4 \pi  P r_{B}^3+r_{B}\right)^2}-4 \pi  P r_{B}^2 \left(2 \pi  P r_{B}^2+1\right).
	\label{TRP}
\end{gather}
	
\begin{figure}	
	\centering
	\includegraphics[scale=1.0]{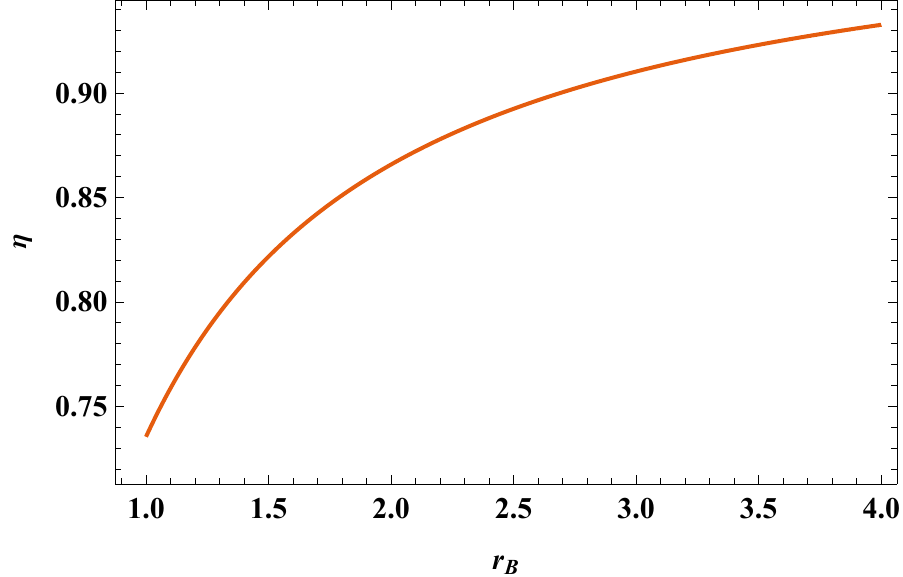}
	\caption{The Carnot cycle efficiency of Schwarzchild black hole heat engine with varying radius $R$, where we set $r_{B1}=1$, $P_{1}=1$, $P_{4}=4$.}
	\label{eta1}
\end{figure}

We denote the heat absorbed as $Q_{H}$, and the heat delivered as $Q_{C}$, so that the mechanical work is $W = Q_{H}-Q_{C}$. The efficiency is the ratio of mechanical work $W$ to heat absorbed $Q_{H}$ %\cite{Hendi:2017bys},
\begin{eqnarray}
	\eta &=& \frac{W}{Q_{H}} = 1 - \frac{Q_{C}}{Q_{H}}. \label{efficiency}
\end{eqnarray} 
The Carnot cycle consists of two isothermal paths and two adiabatic paths which means that the Carnot heat engine is between two different temperatures. We define the high temperature as $T_{H}(r_{B2}, P_{2})$ and the cold one as $T_{L}(r_{B1}, P_{1})$, which are connected through the isochoric paths. The efficiency yields as
\begin{eqnarray}
	\begin{aligned}
		\eta=1-\frac{T_{L}(r_{B1}, P_{1})}{T_{H}(r_{B4}, P_{4})},\\
		%=\frac{R_{4} \sqrt{16 \pi  P_{4} R_{4}^2 \left(2 \pi  P_{4} R_{4}^2+1\right)-4 \sqrt{2 \pi } \sqrt{P_{4} \left(2 \pi  P_{4} R_{4}^2+1\right) \left(4 \pi  P_{4} R_{4}^3+R_{4}\right)^2}+1} \left(\sqrt{2 \pi } \sqrt{P_{4} \left(2 \pi  P_{4} R_{4}^2+1\right) \left(4 \pi  P_{4} R_{4}^3+R_{4}\right)^2}-4 \pi  P_{4} R_{4}^2 \left(2 \pi  P_{4} R_{4}^2+1\right)\right)}{\sqrt{16 \pi  P_{1} R_{1}^2 \left(2 \pi  P_{1} R_{1}^2+1\right)-4 \sqrt{2 \pi } \sqrt{P_{1} \left(2 \pi  P_{1} R_{1}^2+1\right) \left(4 \pi  P_{1} R_{1}^3+R_{1}\right)^2}+1} \left(4 \pi  P_{1} R_{1}^3 \left(2 \pi  P_{1} R_{1}^2+1\right)-\sqrt{2 \pi } R_{1} \sqrt{P_{1} \left(2 \pi  P_{1} R_{1}^2+1\right) \left(4 \pi  P_{1} R_{1}^3+R_{1}\right)^2}\right)}+1
	\end{aligned}
\end{eqnarray}
by which we plot how it varies with respect to the radius of a cavity in \figref{eta1}.

Moreover, we define a rectangular cycle with two isochoric and two isobaric paths and compare its efficiency with that of Carnot cycle, and the efficiency for the rectangle cycle can be calculated by a formula deduced in \cite{Johnson:2016pfa, Rosso:2018acz}
\begin{eqnarray}
	\eta=\frac{W}{Q}=\frac{(P_{1}-P_{4})(V_{4}-V_{1})}{Q},
	\label{eta}
\end{eqnarray}
where
\begin{eqnarray}
	\begin{aligned}
		Q&=\Delta E+ P\Delta V.%\\
		%&= E_{4}-E_{1}+\frac{4\pi}{3}(R_{4}^{3}-R_{4}^{3}).
	\end{aligned}
	\label{H}
\end{eqnarray}
Thus the \myref{eta} yield as
\begin{eqnarray}
	\eta=\frac{4 \pi  (P_{1}-P_{4}) \left(r_{B1}^3-r_{B4}^3\right)}{r_{B4} \left(-4 \pi  P_{1} r_{B4}^2+3 \Delta_{4} -3\right)+4 \pi  P_{1} r_{B1}^3-3 r_{B1} \left(\Delta_{1}-1\right)},
\end{eqnarray}
where we define \[
\Delta_{\alpha}=\sqrt{16 \pi  P_{\alpha} r_{B\alpha}^2 \left(2 \pi  P_{\alpha} r_{B\alpha}^2+1\right)-4 \sqrt{2 \pi } \sqrt{P_{\alpha} \left(2 \pi  P_{\alpha} r_{B\alpha}^2+1\right) \left(4 \pi  P_{\alpha} r_{B\alpha}^3+r_{B\alpha}\right)^2}+1}	
.\] and $\alpha=1, 4$.

In this way, we can show how the efficiency changes as well. 
\begin{figure}	
	\centering
	\includegraphics[scale=1.0]{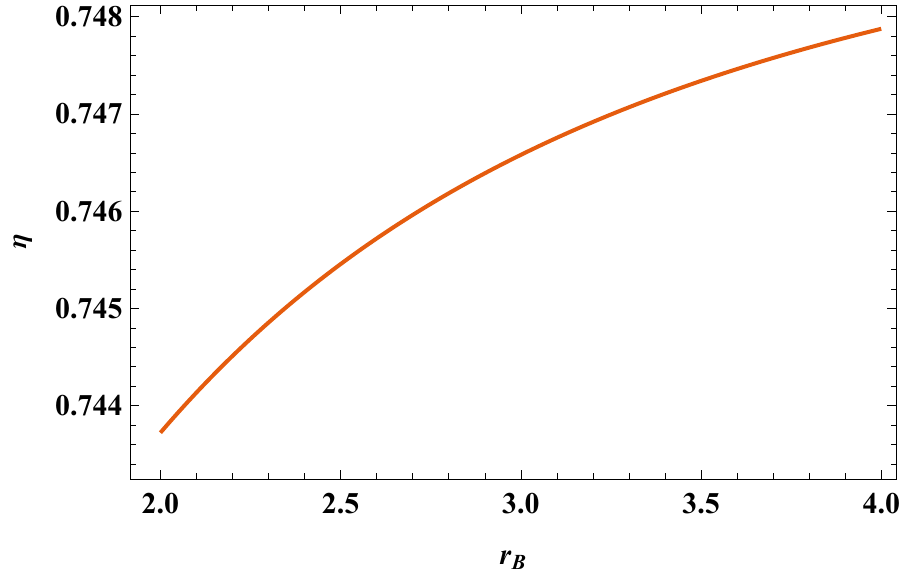}
	\caption{The rectangular cycle efficiency in the Schwarzchild case with varying radius $R$, where we set $r_{B1}=1$, $P_{1}=1$, $P_{4}=4$}
	\label{eta2}
\end{figure}
\subsection{RN black holes in a cavity}
\label{sec:RN}
Then we discuss a more complicated situation. Different from the case of Schwarzchild, we can't analytically derive the expression of $x$ in terms of $r_{B}$ and $P$. Alternatively, we calculate the efficiency numerically. 

In the 4-dimension, the conjugate thermodynamic pressure is
\begin{equation}
	P(r_{B},x,Q)=-\frac{\partial E}{\partial V}=\frac{2r_{B}^{2}x-Q^{2}-r_{B}^{2}x^{2}}{8\pi r_{B}^{4}x\sqrt{f(r_{B})}}-\frac{1}{4\pi r_{B}^{2}}.
	\label{P}
\end{equation}
For every pressure we have fixed, we can solve the corresponding $x$ by \myref{P} with conditions $0<x\leq 1$ and $x>\frac{Q^{2}}{R^{2}}$	which guarantee $P$ and $T$ are real,
and then we can obtain the temperature with $x$ to derive the efficiency of Carnot cycle. Similarly, we can plot the efficiency curve in the rectangular cycle.  		
\begin{figure}[htbp]	
	\centering
	\subfigure[]{\begin{minipage}{8cm}
			\includegraphics[width=0.9\linewidth]{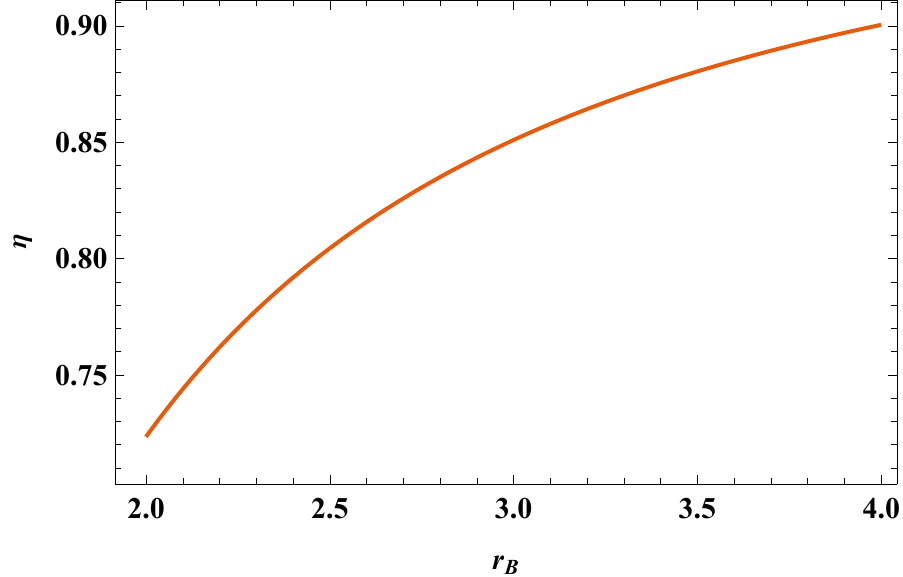}
	\end{minipage}}
	\subfigure[]{\begin{minipage}{8cm}
			\includegraphics[width=0.9\linewidth]{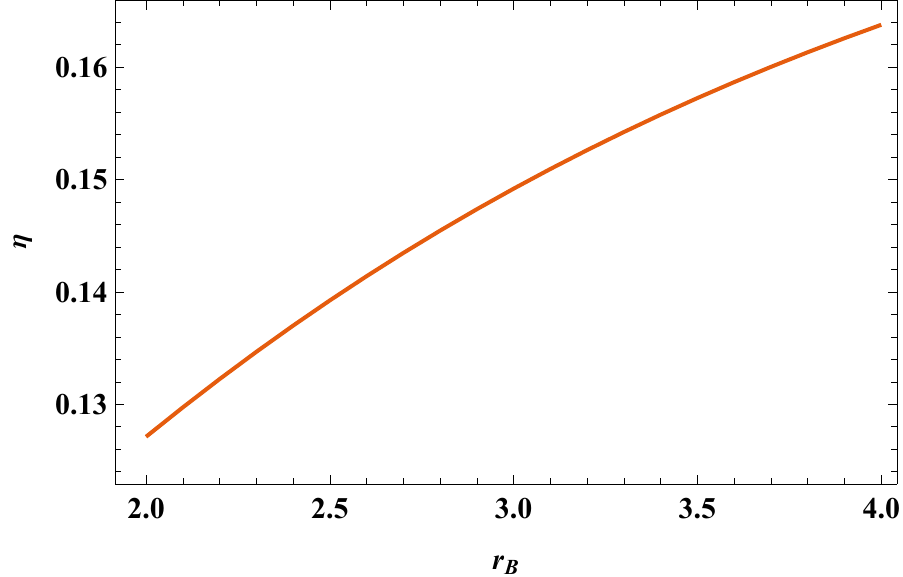}
	\end{minipage}}
	\caption{Left Plane: The Carnot cycle efficiency of RN black hole heat engine; Right Plane: The rectangular cycle efficiency of RN black hole heat engine. Here we set $r_{B1}=1$, $P_{1}=0.01$, $P_{4}=0.02$}
\end{figure}

\section{The JT expansion of RN black hole in a cavity}

\label{sec:JT}

The JT expansion remains a characteristic of the Van der Waals system and the system undergoes an isenthalpic process. As the Joule-Thomson expansion takes place in an isenthalpic process, the enthalpy has an important role in the issues. 

\begin{figure}[H]
	\centering
	\includegraphics[width=.55\linewidth]{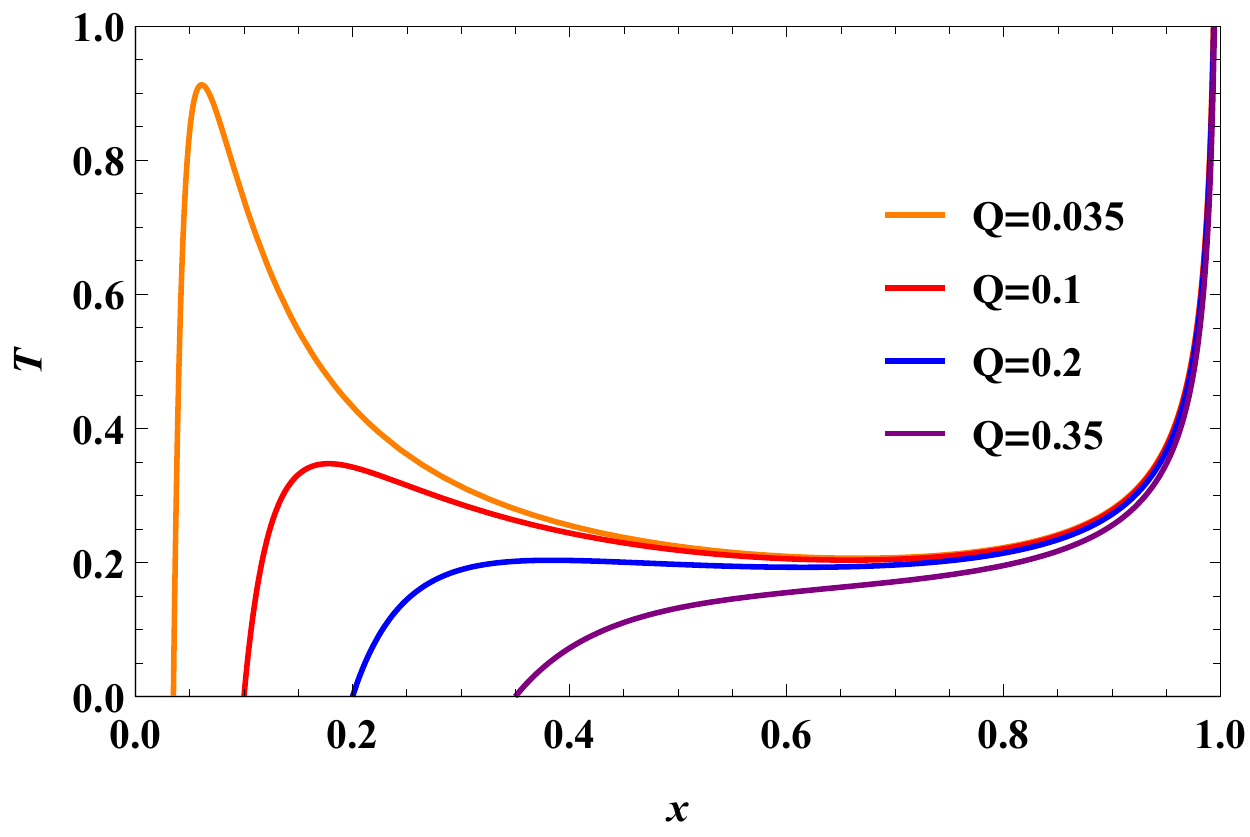}
	\label{fig:tx}
	\caption{We set $Q=0.035,0.1,0.2$ and $0.35$ from top to the bottom. And $r_{B}=1$
		while $x$ have a range of $[0,1]$ in $T-x$ plane.}
	\label{fig:T}
\end{figure}

First of all, four dimensional temperature graph in FIG.\ref{fig:T} according to Eq.(\ref{eq:Th}) shows how the temperature $T$ varies with respect to $x$. It is similar to the temperature graph of RN-AdS black holes \cite{Chamblin:1999hg} only with one difference, that is when $x$ approaches $1$, the temperature diverges. That will be unexpected to have a infinitely high temperature, so we would like to investigate with lower value of $x$.
For further study, we derive the relating thermodynamic quantities. 
The thermal energy in this system have been discussed in \cite{Carlip:2003ne}
and from Eq.\eqref{eq:RNf-1-1}, we obtain the equations of the thermal energy of the system, which yields
\begin{equation}
	E(r_{B},x,Q)=r_{B}(1-\sqrt{f(r_{B})}).\label{eq:E}
\end{equation}
Moreover, the volume of the system is defined to be related to the radius of a cavity, which is 
\begin{equation}
	V=\frac{4}{3}\pi r_{B}^{3}\label{eq:V},
\end{equation}
whose conjugate thermodynamic pressure naturally arises as
\begin{equation}
	P(r_{B},x,Q)=-\frac{\partial E}{\partial V}=\frac{2r_{B}^{2}x-Q^{2}-r_{B}^{2}x^{2}}{8\pi r_{B}^{4}x\sqrt{f(r_{B})}}-\frac{1}{4\pi r_{B}^{2}}.
	\label{eq:P4}
\end{equation}
With the three equations above \eqref{eq:E}, \eqref{eq:V} and \eqref{eq:P4}, the enthalpy of the system is obtained by employing the relation $H=E+PV$, 
\begin{eqnarray}
	H(r_{B},x,Q) & = & \frac{-r_{B}^{2}x\left(6f(r_{B})-4\sqrt{f(r_{B})}+x-2\right)-Q^{2}}{6\sqrt{f(r_{B})}r_{B}x}\label{eq:H4}.
\end{eqnarray}

\begin{figure}[H]
  	\centering
  	\subfigure[]{\begin{minipage}{8cm}
 	\includegraphics[width=0.9\linewidth]{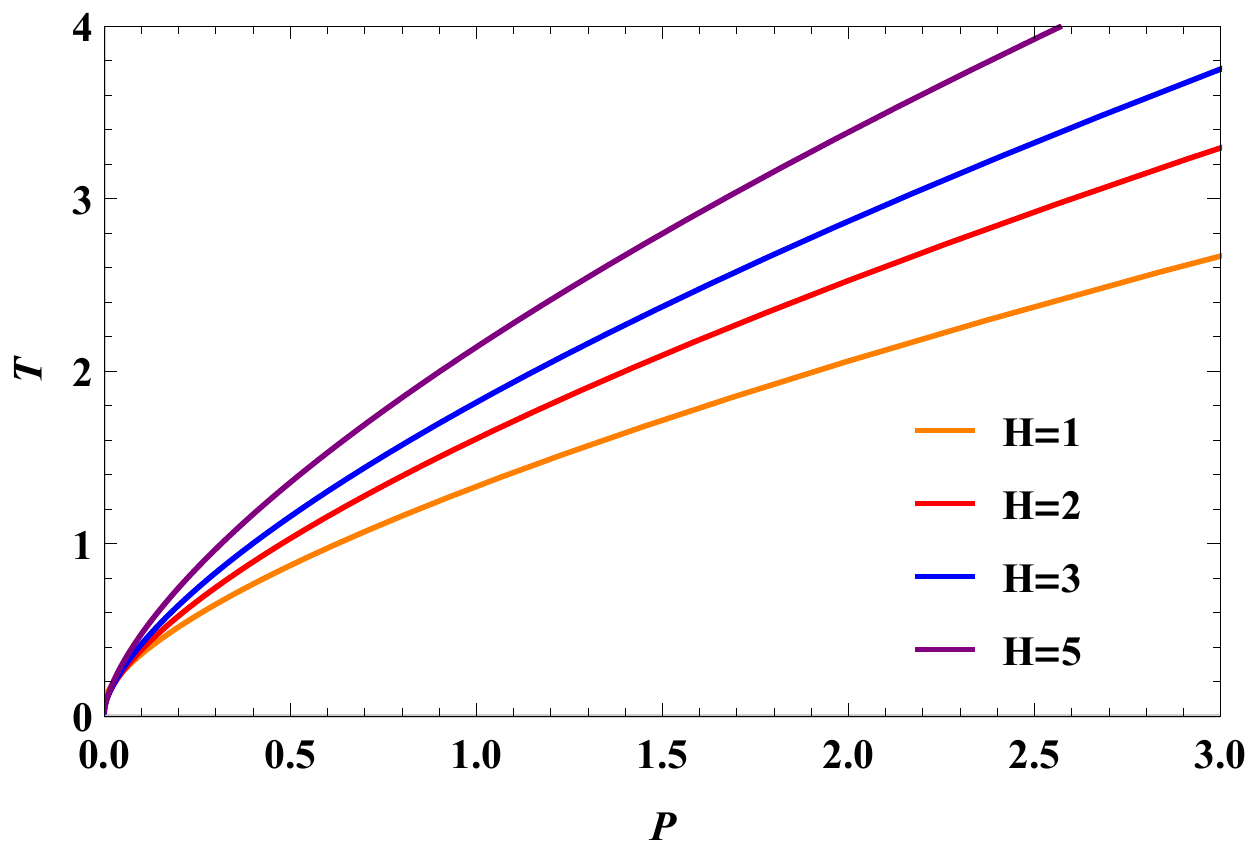}
 	 \label{d4high}
 \end{minipage}}
\subfigure[]{\begin{minipage}{8cm}
	\includegraphics[width=0.94\linewidth]{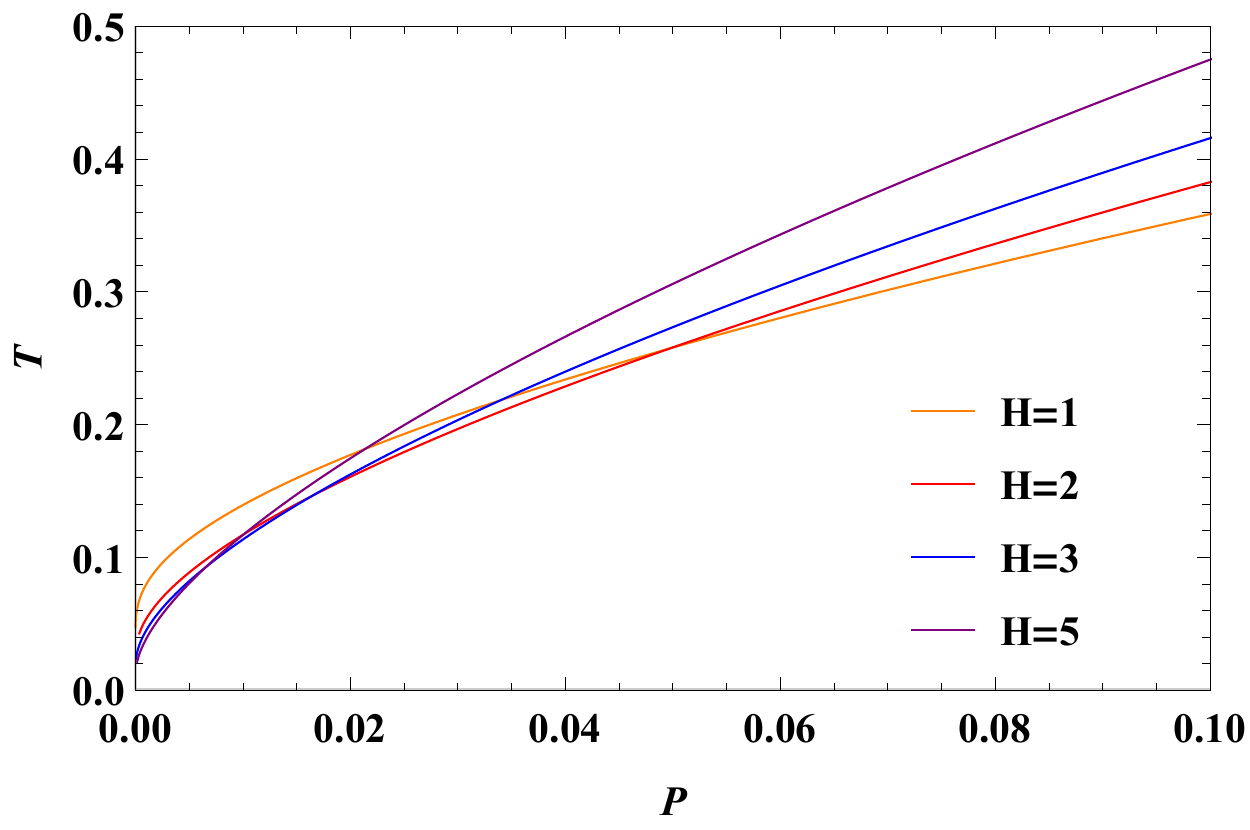}
 \label{d4low}
\end{minipage}}
 	\caption{Isenthalpic curves in $T-P$ plane. $H=1,2,3$ and $5$. $Q=0.035$, $d=4$, and $r_{B}$ varies from $0.2$ to $10$.}
 	\label{fig:tpd4h1235}
 \end{figure}

We have obtained all the quantities required and are able to study the JT expansion of RN black hole in a cavity. The isenthalpic curves can be obtained from Eqs.\eqref{eq:T4}, \eqref{eq:P4} and \eqref{eq:H4} numerically in FIG. \ref{fig:tpd4h1235}.

\begin{figure}[H]
	\centering
	\includegraphics[width=0.55\linewidth]{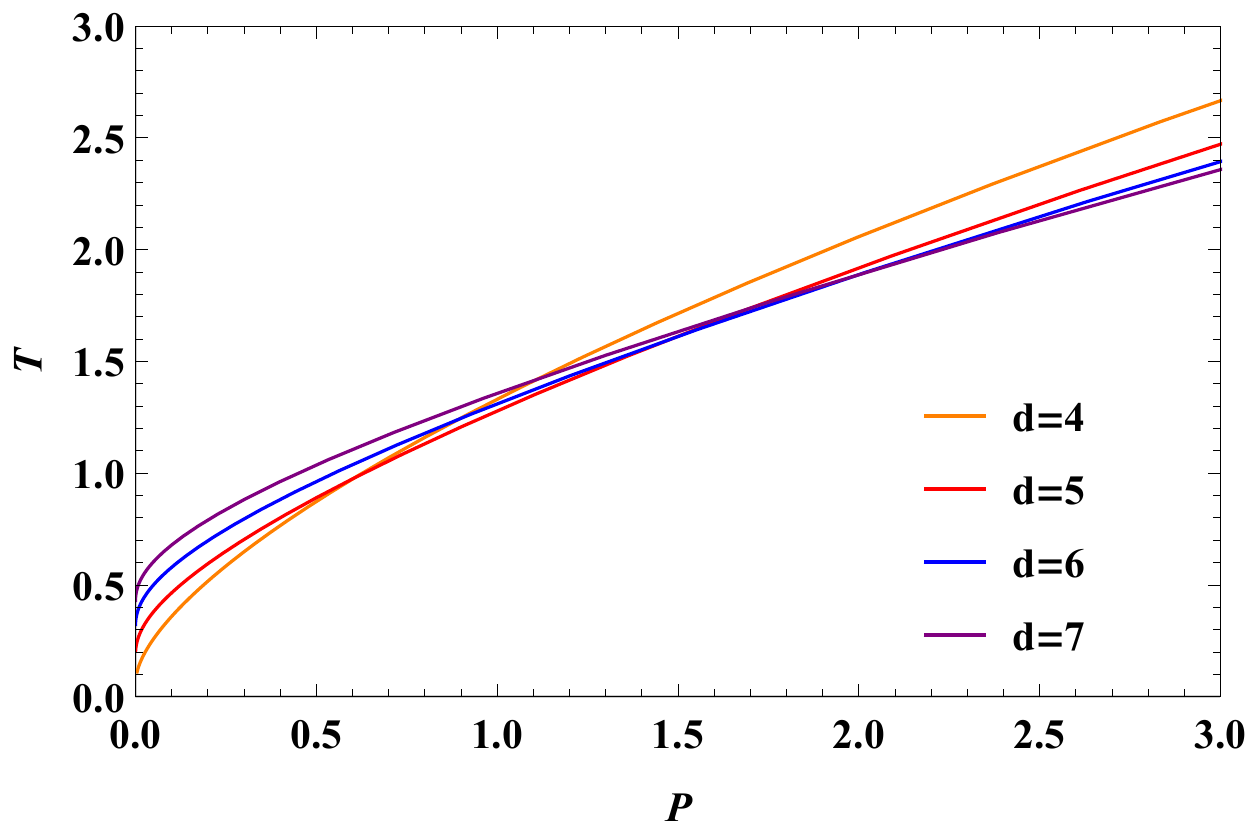}
	\caption{The isenthalpic curves of charged black holes in a cavity for high dimensional cases. We set the dimension $d$ equals $4,5,6$ and $7$ respectively, the enthalpy $H$ equals $1$, and the charge $Q$ equals $0.035$.}
	\label{fig:tpd4567}
\end{figure}

As we can see, there are no extreme points in the curves and the temperature decreases with lower pressure. Furthermore, when the pressure and temperature are high, the lines from bottom to the top show the isenthalpic curves of increasing value of the enthalpy, $H=1,2,3$ and $5$, which is obvious in FIG.\ref{d4high}. Moreover, in FIG.\ref{d4low}, we show the situation of lower temperature and pressure. The part of the curves that decreases as the enthalpy rise corresponds to smaller $x$, that is, when the ratio of event horizon $r_{+}$ and cavity radius $r_{B}$ is small, and it is in line with the range we want when we study the temperature curves. And the orange line coincide with the orange one in FIG.\ref{fig:tpd4567} which represent the four dimensional cases when the enthalpy equals $1$. 

We can investigate the project further. For d-dimensional cases, the explicit form of $P(r_{+},r_{B},q,d)$ and $H(r_{+},r_{B},q,d)$ takes the form as
\begin{equation}
	P=-\frac{\partial E}{\partial V}=\frac{\left(d-2\right)(d-3)}{8\pi}\Xi,\label{eq:Ph}
\end{equation}
\[
\Xi=-\frac{q^{2}}{2\sqrt{f(r_{B})}r_{B}^{3d-6}\text{\ensuremath{r_{+}}}^{d-3}}-\frac{1}{2\sqrt{f(r_{B})}\text{\ensuremath{r_{B}^{d-1}r_{+}}}}+\frac{1}{\sqrt{f(r_{B})}r_{B}^{2}}-\frac{1}{r_{B}^{2}}.
\]

\begin{figure}[H]
	\centering 
	\includegraphics[width=0.55\linewidth]{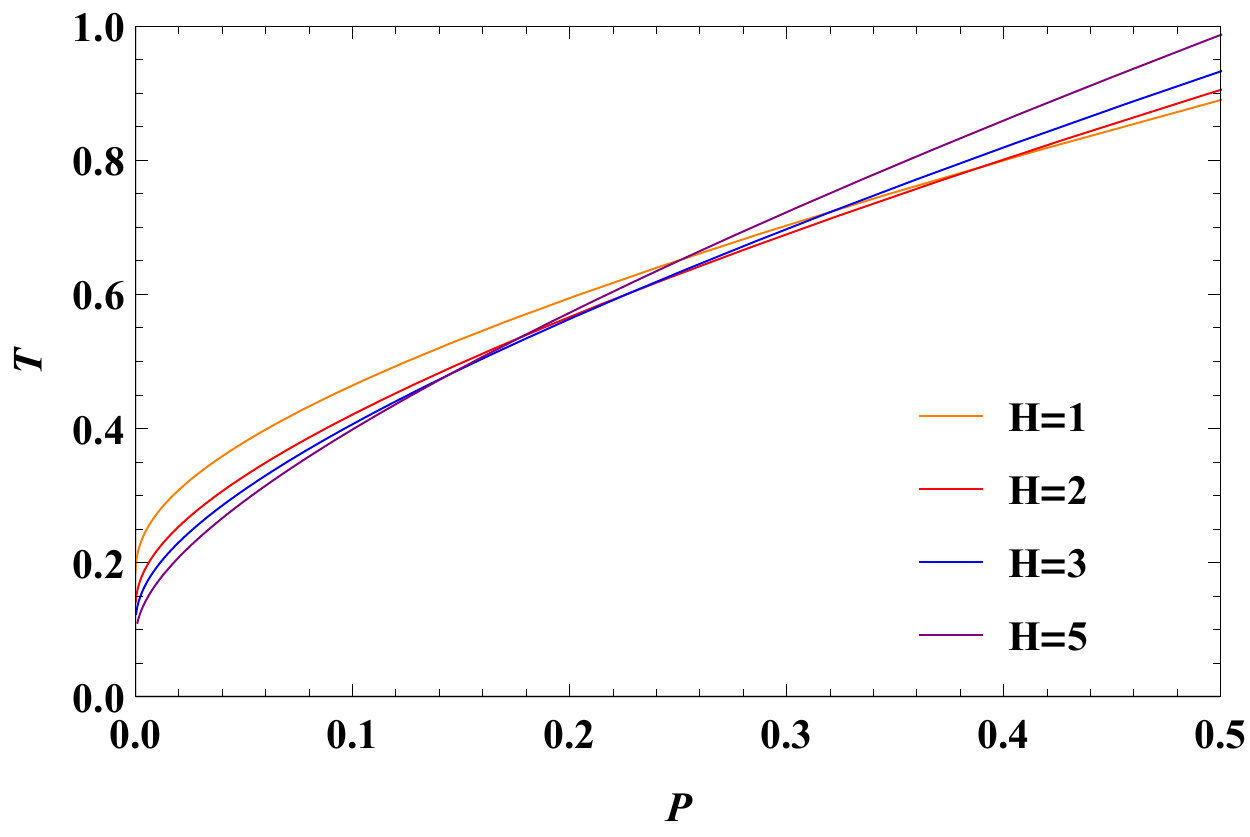}
	\caption{The isenthalpic curves of charged black hole in cavity. $H=1,2,3$ and $5$. $d=5$, $Q=0.035$, and $r_{B}$ varies from $0.2$ to $3$.}	
	\label{fig:tpd5h12345}
\end{figure}

The enthalpy $H$ is derived by $H=E+PV$, which yields
\begin{equation}
	H=\frac{(d-2)r_{B}^{d-3}\omega_{d-2}}{8\pi}(1-\sqrt{f(r_{B})}+\frac{(d-3)r_{B}^{2}}{d-1}\Xi).\label{eq:H}
\end{equation}
We obtain the isenthalpic curves in numerically way with Eqs.\eqref{eq:Th}, \eqref{eq:Ph} and \eqref{eq:H}, and then we plot how the enthalpy varies from $1$ to $5$ when $d=5, Q=0.035$ in FIG.\ref{fig:tpd5h12345}. 

The temperature decreases with pressure as shown in FIG.\ref{fig:tpd5h12345}, which shows that the process only cools the system. The graph can be divided with the curves and the left part represents the low pressure region, while the right part indicates the high pressure region. Additionally, from bottom to the top, the temperature in low pressure region increases with decreasing enthalpy and increases with the increasing enthalpy in the high pressure region. Similar to the four-dimensional case, in the low pressure region, the ratio of $r_{+}$ and $r_{B}$ is small.

After studying the four and five dimensional cases, we would like to consider the impact of dimensions on the isenthalpic curves.

The picture FIG.\ref{fig:tpd4567} presents different dimensional cases where the red and the orange curve coincides with the orange one in FIG.\ref{fig:tpd5h12345} and the orange one in FIG.\ref{fig:tpd4h1235} respectively. We can see from the picture that, in the low pressure area, higher dimensions lead to higher temperatures, which is the opposite in the high pressure region. The influence of the dimension parameter $d$ and the enthalpy $H$ on the temperature during the throttling process is opposite. 

\begin{figure}[H]
	\centering
	\includegraphics[width=0.45\linewidth]{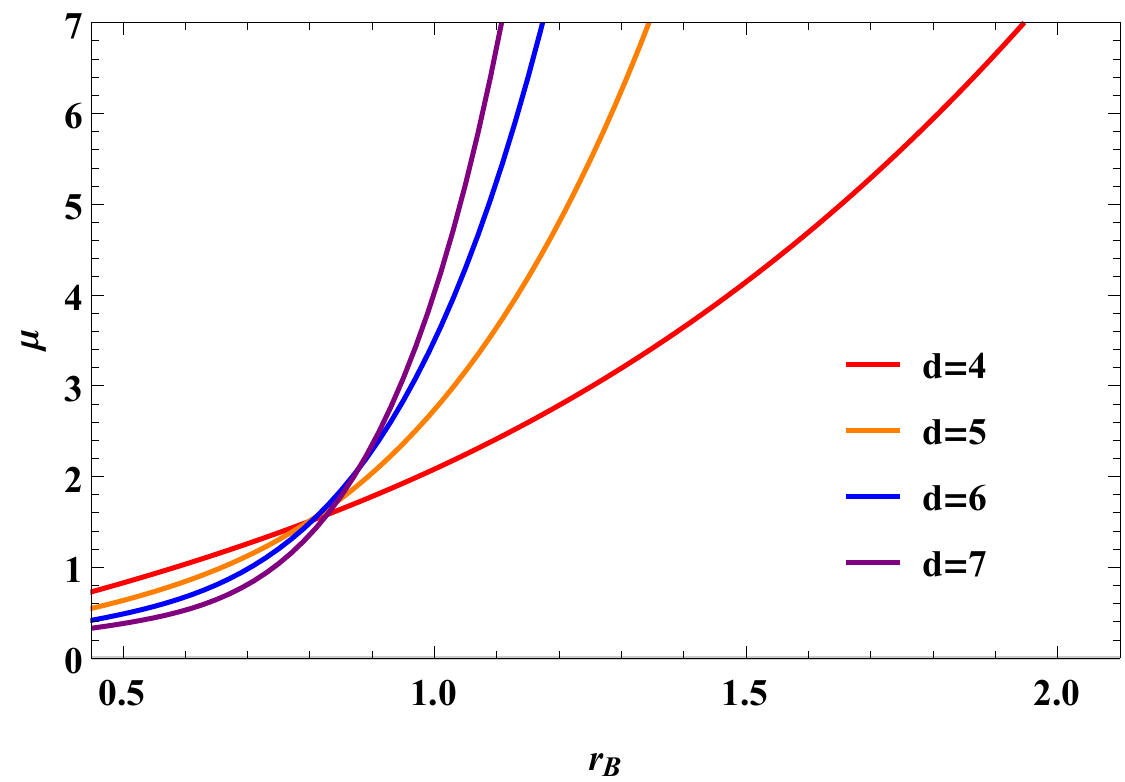}	
	\caption{The Joule-Thomson coefficient of the charged RN black hole in a cavity for $d=4,5,6$ and $7$. The enthalpy $H=1$, $r_{B}$ varies from $0.5$ to $2.5$, and ${Q}=0.035$.}	
	\label{fig:u4567r}
\end{figure}

As we know, the JT coefficient of classical ideal gas always equals $0$, which means the isnethalpic process doesn't change the temperature of ideal gas, while the process can cool or heat the classical van der Waals gas only depending on the initial temperature and pressure. With Eqs.\eqref{eq:T4} and \eqref{eq:P4}, we derive JT coefficient for 4 dimensional case by
 \begin{equation}
 \mu=(\frac{\partial T}{\partial P})_{H},
 \end{equation}
  and show how it changes with $r_{B}$ in FIG.\ref{fig:u4567r} with the red line. The JT coefficient $\mu$ obtained numerically with Eqs.\eqref{eq:Th} and \eqref{eq:Ph} in different dimensions are shown in FIG.\ref{fig:u4567r} too. When $\mu=(\frac{\partial T}{\partial P})_{H}=0$, the transition between cooling and heating region occurs. But for d-dimensional charged black holes in a cavity, the Joule-Thomson coefficient is positive infinite.

 We conclude Joule-Thomson expansion is characterized by the invariance of enthalpy and Joule-Thomson coefficient $\mu$ determines the final change of temperature during the process. One can use the sign of $\mu$ to divide the heating and cooling zone. We can see from FIG.\ref{fig:u4567r} that $\mu$ is always positive, which means no transition occurs. Comparing the isenthalpic curves above with the JT coefficient graph, the property of positive definite and the constant phase of cooling are obvious.

\section{Conclusions}
 Based on the thermodynamics of black holes in a cavity, we derive the temperature, pressure and other thermodynamic quantities in d-dimension to discuss the heat engine and the	Joule-Thomson expansion of charged black holes. As a cavity defines an explicit boundary for the working substance, i.e. black holes, it's more rational to consider a heat engine in cavities rather than in AdS space, whose scale corresponds to the pressure in the extended space. Therefore, we derive the efficiency of Schwarzchild and RN black hole heat engines in the Carnot cycle and rectangular cycle and show how it changes with respect to the radius of a cavity. Moreover, we find the Joule-Thomson expansion of charged black holes exist in a cavity but differ from the AdS cases. We conclude the black hole always cools down during the isenthalpic process with the decreasing pressure, while the JT coefficient $\mu$ keeps positive infinite.

\label{sec:Discussion-and-Conclusions}
\begin{acknowledgments}
We are grateful to Yucheng Huang and Peng Wang for useful discussions and valuable comments. This work is supported by NSFC (Grant No.11947408 and 12047573).
\end{acknowledgments}

\bibliographystyle{unsrturl}
\nocite{*}
\normalem
\bibliography{JTcavity.4.0}

\end{document}